\documentclass[prl,twocolumn,showpacs,byrevtex,amsmath,amssymb]{revtex4}
\usepackage{graphicx}
\usepackage{dcolumn}
\usepackage{bm}

\begin{document}
\preprint{PREPRINT}

\title[Short Title]{Reentrant transitions in colloidal or dusty plasma bilayers}

\author{Ren\'e Messina}
\email{messina@thphy.uni-duesseldorf.de}
\affiliation{
 Institut f\"ur Theoretische Physik II, 
 Heinrich-Heine-Universit\"at D\"{u}sseldorf,
 Universit\"{a}tsstrasse 1,
 D-40225 D\"{u}sseldorf, Germany }
\author{Hartmut L\"owen}
\email{hlowen@thphy.uni-duesseldorf.de}
\affiliation{
 Institut f\"ur Theoretische Physik II, 
 Heinrich-Heine-Universit\"at D\"{u}sseldorf,
 Universit\"{a}tsstrasse 1,
 D-40225 D\"{u}sseldorf, Germany }

\date{\today}

\begin{abstract}
The phase diagram of crystalline  bilayers of particles interacting via a Yukawa
potential is calculated for arbitrary screening lengths and particle densities.
Staggered rectangular, square, rhombic and triangular structures are found to
be stable including  a first-order transition between two different rhombic structures. 
For varied screening length at fixed density, one of these rhombic phases exhibits 
both a single and even a double reentrant transition.
Our predictions can be verified
experimentally in strongly confined charged colloidal suspensions 
or dusty plasma bilayers.
\end{abstract}
%
%
%
%
\maketitle

Confined systems exhibit structural and dynamical behavior 
very much different from the corresponding
 bulk state \cite{Binderreview,Bechingerreview}. 
In particular freezing is strongly affected by the presence of a 
planar wall. In equilibrium, solidification near  walls
can occur at thermodynamic conditions where the bulk
is still fluid  (so called ``prefreezing'')
\cite{vanSwol,Heni2}. In nonequilibrium, the wall may act as a 
center of heterogeneous nucleation \cite{Palberg1}
in order to initiate crystal growth \cite{vanBlaaderen}.
A system confined between two parallel planar walls 
exhibits various layered crystalline states at low temperature
if the plate distance gets comparable to the mean interparticle distance. 
For hard spheres between hard plates,
geometric packing considerations lead to the stability 
of different crystalline lattices including multiple square and hexagonal 
layers \cite{Murray} as well as buckled \cite{Pansu}, rhombic 
\cite{Pansu,Schmidt} and prism superlattices \cite{Bechinger}. 
On the other hand, for pure coulombic systems
like (classical) electrons in quantum wells 
\cite{Kalman} or trapped ions \cite{Raizen},
several crystalline bilayer-structures were reported \cite{Peeters}.

Most of our experimental
knowledge of freezing in confining slit-like geometry is based on real-space
measurements  of mesoscopic model systems
such as charged colloidal suspensions between glass plates \cite{Murray,Bechinger} or 
of multilayers of highly charged dusty plasmas \cite{plasma2}. The actual
interaction between these mesoscopic ``macroions'' is neither hard-sphere like
nor pure coulombic but is described by an intermediate screened Coulomb or Yukawa
pair potential \cite{Robbins,test} due to the screening via additional microions 
in the system. The screening length can be tailored by changing the  microion concentration:
for charged colloids, salt ions are conveniently  added to the aqueous suspensions;
the complex plasma, on the other hand,  consists of electrons and impurity ions.

In the present letter, we study the stability
of different crystal lattices in bilayers of Yukawa particles as motivated by the 
experimental model systems. The zero-temperature phase diagram is calculated
for arbitrary screening lengths and particle densities \cite{note_peeters}.
We find a variety of different staggered solid lattices
to be stable which are separated by either first or second-order phase 
transitions. The two known extreme limits of zero
or infinite screening length corresponding to  hard-spheres \cite{Schmidt}
and the plasma \cite{Peeters,Goldoni} are recovered.
For intermediate screening lengths, the phase
behavior is strikingly different from a simple interpolation between these two limits.
First, there is a first-order coexistence between two different staggered rhombic lattices
differing in their relative shift of the two unit cells.
Second, one of these staggered rhombic phases exhibits a novel
reentrant effect for fixed density and varied screening length. Depending on the
density, the reentrant transition can proceed via  a staggered square or a
staggered triangular solid including even a {\it double reentrant transition} of the rhombic phase.
All our theoretical predictions can in principle be verified in real-space 
studies of confined  charged suspensions or dusty plasmas.

In detail, our system consists of two layers containing in total $N$
particles in the $(x,y)$ plane. The total area density of the two layers is
$\rho= N/A$ with $A$ denoting the system area in the $(x,y)$ plane. The distance $D$ between
the layers in the $z$ direction is prescribed by the external potential confining the system.
The  particles are interacting via the Yukawa pair potential 
%
\begin{equation}
\label{Eq.Yukawa}
V(r) =  V_0 \frac{\exp(-\kappa r)}{\kappa r},
\end{equation}
%
where $r$ is the central separation. The inverse screening length
$\kappa$  which governs the range of the interaction is given in terms
of the micro-ion concentration by Debye-H\"uckel
screening theory. The energy amplitude $V_0 = Z^2\exp (2\kappa R)\kappa/\epsilon
(1+\kappa R)^2$ scales with the square of the charges $Z$ of the particles
of physical hard core radius  $R$ \cite{point_particle} reduced by the dielectric 
$\epsilon$ permittivity of the solvent ($\epsilon=1$ for the dusty plasma).
Typically, $Z$ is of the order of $100-100000$ elementary charges
such that $V(r)$ at typical interparticle distances can be much larger than the thermal 
energy $k_BT$ at room temperature
justifying formally zero-temperature calculations. Then the energy scale
is set by $V_0$ alone and  phase transitions in large bilayer systems
are  completely determined by 
two dimensionless parameters, namely the reduced layer density  $\eta = \rho D^2 /2$ 
and the relative screening strength $\lambda = \kappa D$.
For zero temperature, the stable state is solid  but  different
crystalline structures of the bilayers are conceivable. As possible
candidate structures we assume that the two two-dimensional periodic lattices
in the bilayers are the same, have a simple unit cell and are shifted
relative to each other in the lateral direction by a displacement vector ${\bf c}$.
If the two layers are labeled with $A$ and $B$, the particle 
positions in the $(x,y)$ plane of the two layers 
are given by
%
\begin{eqnarray}
\label{Eq.R_AB}
{\bf R}_A (m,n) & = & m {\bf a_1} + n {\bf a_2}, \nonumber \\
{\bf R}_B (m,n) & = & m {\bf a_1} + n {\bf a_2} + {\bf c},
\end{eqnarray}
%
where ${\bf a_1}$ and ${\bf a_2}$ are the primitive vectors
of the two-dimensional lattice and
$m,n$ are integers. The total internal energy $U$ is  obtained 
by the double lattice sum
%
\begin{eqnarray}
\label{Eq.lattice}
U & = & \frac{1}{2} \sum_{{\bf R}_A \neq {\bf R}'_A} V(|{\bf R}_A - {\bf R}'_A|)
+ \frac{1}{2} \sum_{{\bf R}_B \neq {\bf R}'_B} V(|{\bf R}_B - {\bf R}'_B|) \nonumber \\
&&  + \sum_{{\bf R}_A, {\bf R}_B}  
                                      V([|{\bf R}_A - {\bf R}_B|^2 + D^2]^{1/2}).
\end{eqnarray}
%
In the limit $N \rightarrow \infty$, the stable crystalline structure
minimizes the total internal energy per particle $U/N$. 

%
\begin{table}[b]
\caption{Structure and parameters of the different staggered bilayer crystals. 
${\bf a}_1$ is set to $(a_1,0)$ where $a_1$ is the nearest intralayer distance between particles.
For phase $\text{II}$, $\gamma=a_2/a_1$ is the aspect ratio. 
For phase $\text{IV}$, $\theta$ is the angle between  ${\bf a}_1$ and ${\bf a}_2$, and $\alpha$ 
is a free parameter characterizing the relative lateral interlattice shift ${\bf c}$.}
\label{tab.lattice}
\begin{ruledtabular}
\begin{tabular}{lccc}
%
 Phase  & ${\bf a}_2/a_1$ & ${\bf c}$ & $\rho a_1^2/2$ \\
\hline 
I.   Rectangular     & $(0,\sqrt 3)$    & $({\bf a}_1 + {\bf a}_2)/2$  & $1/\sqrt3$\\
II.  Rectangular     & $(0,\gamma)$    & $({\bf a}_1 + {\bf a}_2)/2$   & $\gamma$\\
III. Square          & $(0,1)$          & $({\bf a}_1 + {\bf a}_2)/2$  & $1$\\
IV.  Rhombic         & $(\cos \theta,\sin \theta)$          
                               & $({\bf a}_1 + {\bf a}_2) \alpha$  & $1/\sin \theta$\\
V.   Triangular      & $(1/2,\sqrt3/2)$ & $({\bf a}_1 + {\bf a}_2)/3$  & $2/\sqrt3$ \\
\end{tabular}
\end{ruledtabular}
\end{table}
%

We have minimized  $U/N$ with respect to ${\bf a_1}$, ${\bf a_2}$ and ${\bf c}$
under the constraint of prescribed density $\eta$ for given $\lambda$
mapping out the phase diagram  in the $( \eta \lambda )$ plane.
As a result, five typical staggered lattice structures turn out to minimize 
$U/N$ for different $\eta$. Adopting the notation
developed for plasma bilayers \cite{Goldoni}, we label them by I, II, III, IV, and V.
As summarized in Table \ref{tab.lattice}, phase I is the staggered rectangular crystal
with a fixed aspect ratio $a_2/a_1$ of $\sqrt{3}$; phase II is also staggered
rectangular but with a different aspect ratio $\gamma$ interpolating continuously between
phase I and the staggered-square phase III where $a_2/a_1=1$. The staggered rhombic phase IV has
two non-orthogonal lattice unit vectors $({\bf a}_1$ and ${\bf a}_2)$ 
forming an angle $\theta$ and contains
a general lateral shift $\bf c=\alpha ({\bf a}_1 + {\bf a}_2) $ between the two rhombic lattices.
In fact, we find  two possibilities for
$\alpha$ defining {\it two variants of stable rhombic phases} which we call 
IVA and IVB. For IVA, $\alpha=1/2$ while
$\alpha<1/2$ for IVB. Finally, phase V is a staggered triangular crystal.
Both phases III and V can be considered as special cases of the rhombic
phase IV; the former has $\theta=\pi/2$  and 
$\alpha = 1/2$ while the latter is characterized by $\theta=\pi/3$  and $\alpha = 1/3$.

\begin{figure}
\includegraphics[width = 8.0 cm]{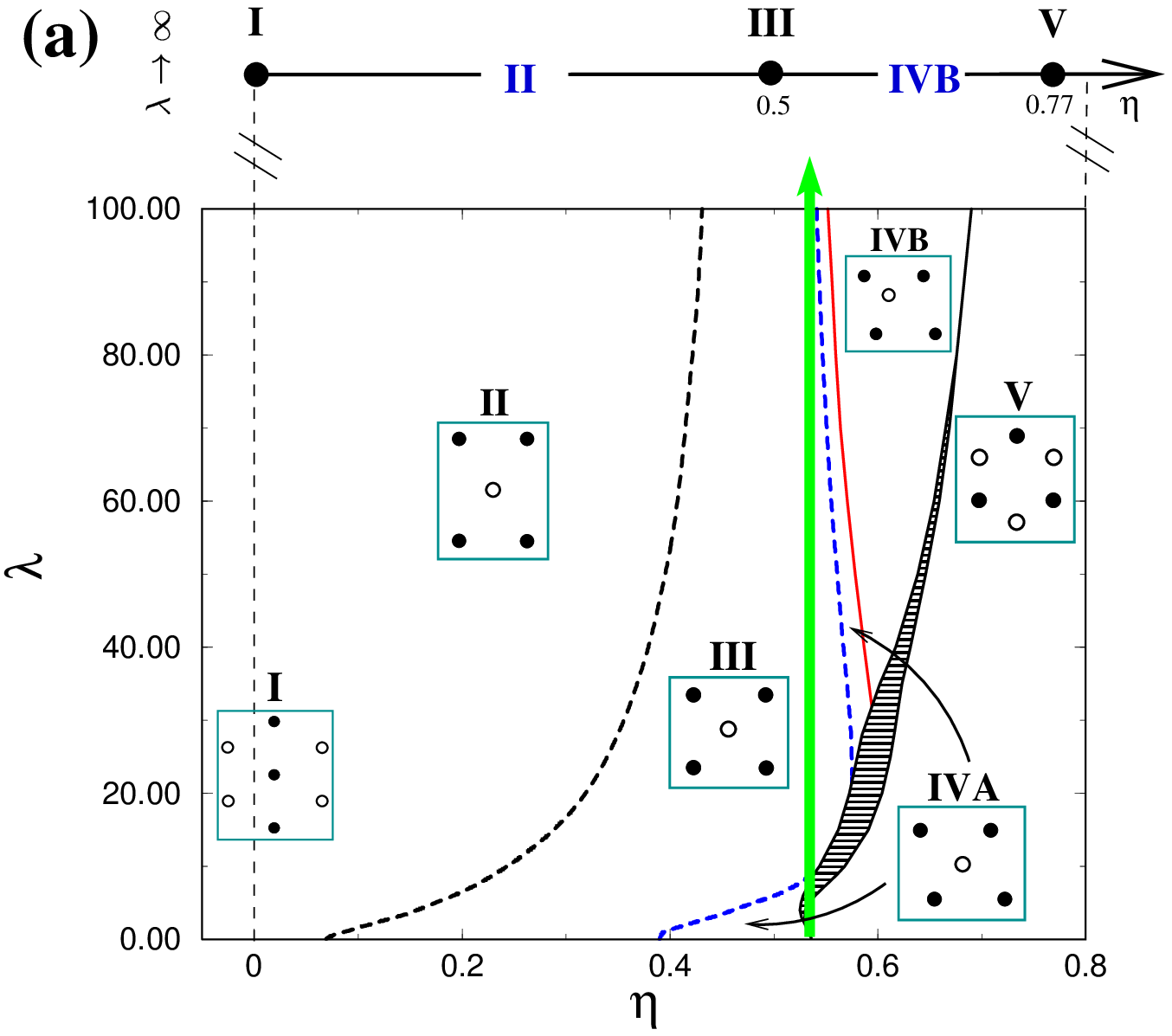}
\includegraphics[width = 8.0 cm]{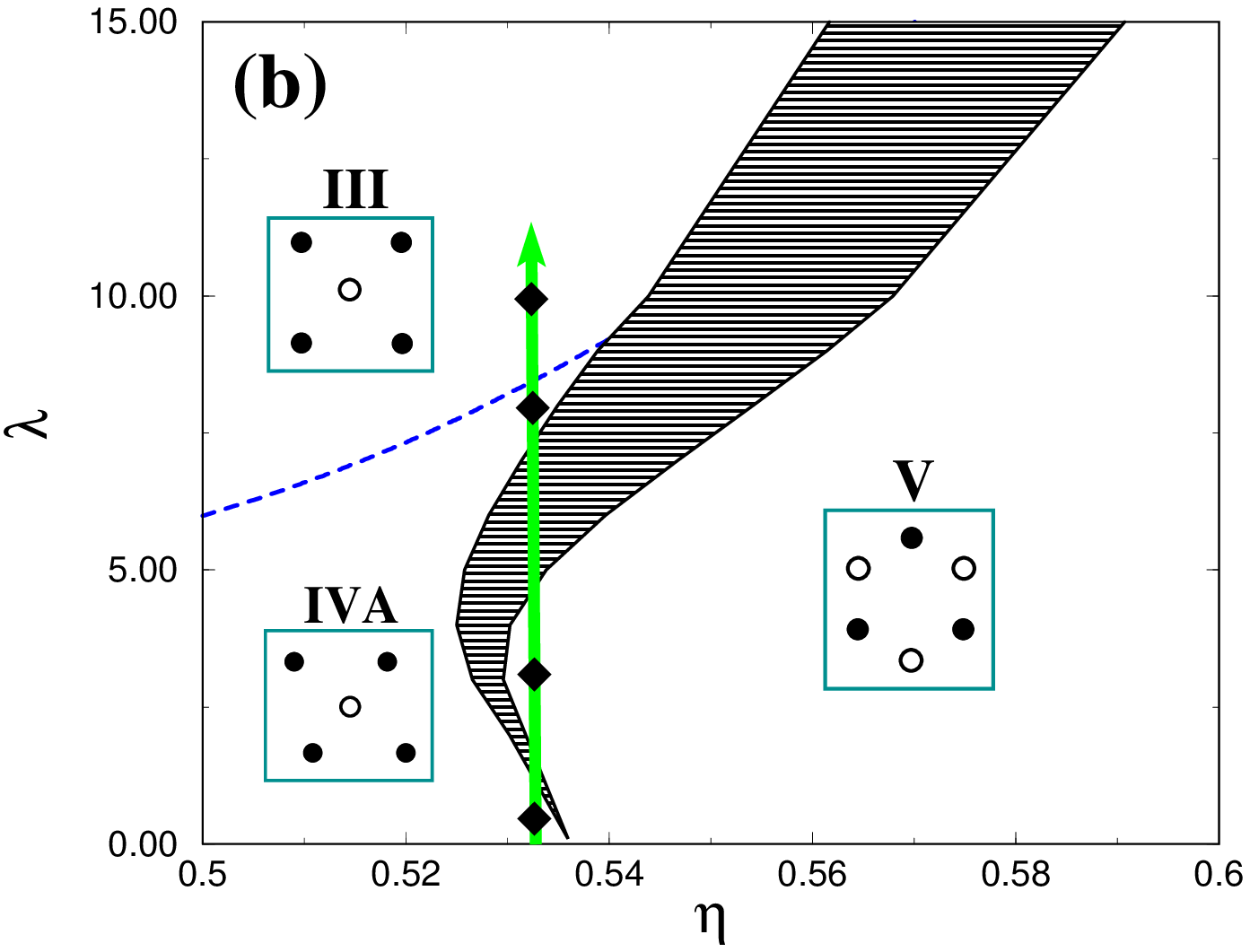}
\caption{Phase diagram of the Yukawa bilayer
in the $(\eta,\lambda)$ plane. 
(a)
The hard sphere limit $\lambda \rightarrow \infty$ is sketched on top.
The dashed (solid) lines denote continuous (discontinuous) transitions. 
The filled region corresponds to the coexistence domain of phases IV and V. 
The vertical arrow indicates the \textit{double} reentrant behavior of phase IVA.
The insets show the lattice geometries,
where the filled (open) circles correspond to the lower (upper) layer.  
(b) Magnification of (a) showing a reentrant behavior of phase IVA
occurring at moderate $\lambda$. The four diamonds along the arrow 
indicate state points which were investigated by computer simulation
at finite temperatures.
}
\label{Fig.diagram} 
\end{figure}

\begin{figure}[b]
\includegraphics[width = 8.0 cm]{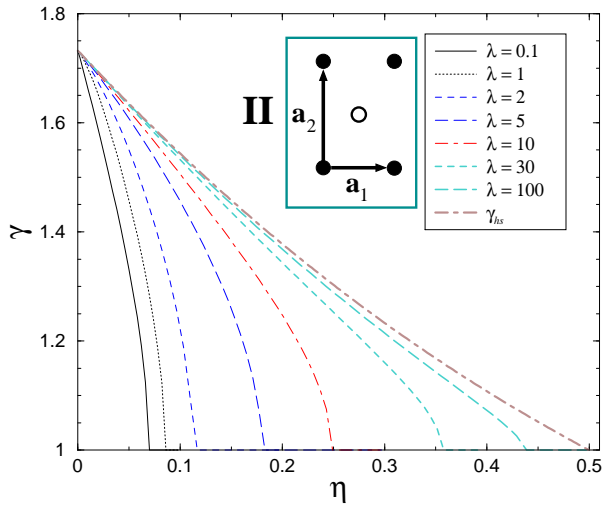}
\caption{Aspect ratio $\gamma=a_2/a_1$  for phase II versus density $\eta$
for different screening strengths $\lambda$. The hard sphere case $\gamma_{hs}$ is also
shown. The lattice geometry is shown as an inset, 
where the filled (open) circles correspond to the lower (upper) layer.
}
\label{fig.gamma}
\end{figure}

The result for the phase diagram  for a wide range of screening strengths 
$(0 \leq \lambda \leq 100; ~\lambda \rightarrow \infty)$ 
and densities $(0 \leq \eta \leq 0.8)$ is shown in Fig. \ref{Fig.diagram}. 
%
%
At very low screening $\lambda$, we recover the known plasma limit \cite{Goldoni}, with
our labeling of the phases being in line with their sequence for increasing density $\eta$.
Phase I has a finite but extremely small density region of stability 
up to $\eta = 3.6 \times 10^{-5}$ at $\lambda=0$ \cite{Goldoni}. 
For finite $\lambda$, the $\text{I} \rightarrow \text{II}$ transition stays second-order and
occurs at even smaller densities which decrease monotonically
to zero as a function of $\lambda$ until the hard-sphere limit $\lambda \to \infty$ is reached.
In this latter case $a_1$ is playing the role of an effective particle diameter.
This is sketched by the vertical line in  Fig. \ref{Fig.diagram}. 
The $\text{II} \rightarrow \text{III}$ transition is second-order and the transition densities
increase drastically  with growing $\lambda$ and interpolating monotonically between
the plasma and hard sphere limit. More details of the 
$\text{I} \rightarrow \text{II} \rightarrow \text{III}$ transition scenario
are depicted in Fig. \ref{fig.gamma} where the aspect ratio $\gamma$ of phase II
is shown versus $\eta$ for different $\lambda$. 
Phases I and III correspond to $\gamma=\sqrt3$ and $\gamma=1$, respectively.
As can be clearly deduced from Fig. \ref{fig.gamma}, the aspect ratio $\gamma$
interpolates continuously as a function of $\eta$ between $\sqrt{3}$ and $1$ for any
$\lambda$ such that both the $\text{I} \rightarrow \text{II}$ and the
$\text{II} \rightarrow \text{III}$ transitions are second-order. In the
hard sphere limit, $\gamma$ approaches $\gamma_{hs} = -2 \eta + \sqrt{4\eta^2+3}$
continuously which is also shown in Fig. \ref{fig.gamma}.

%
%

\begin{figure}
\includegraphics[width = 8.0 cm]{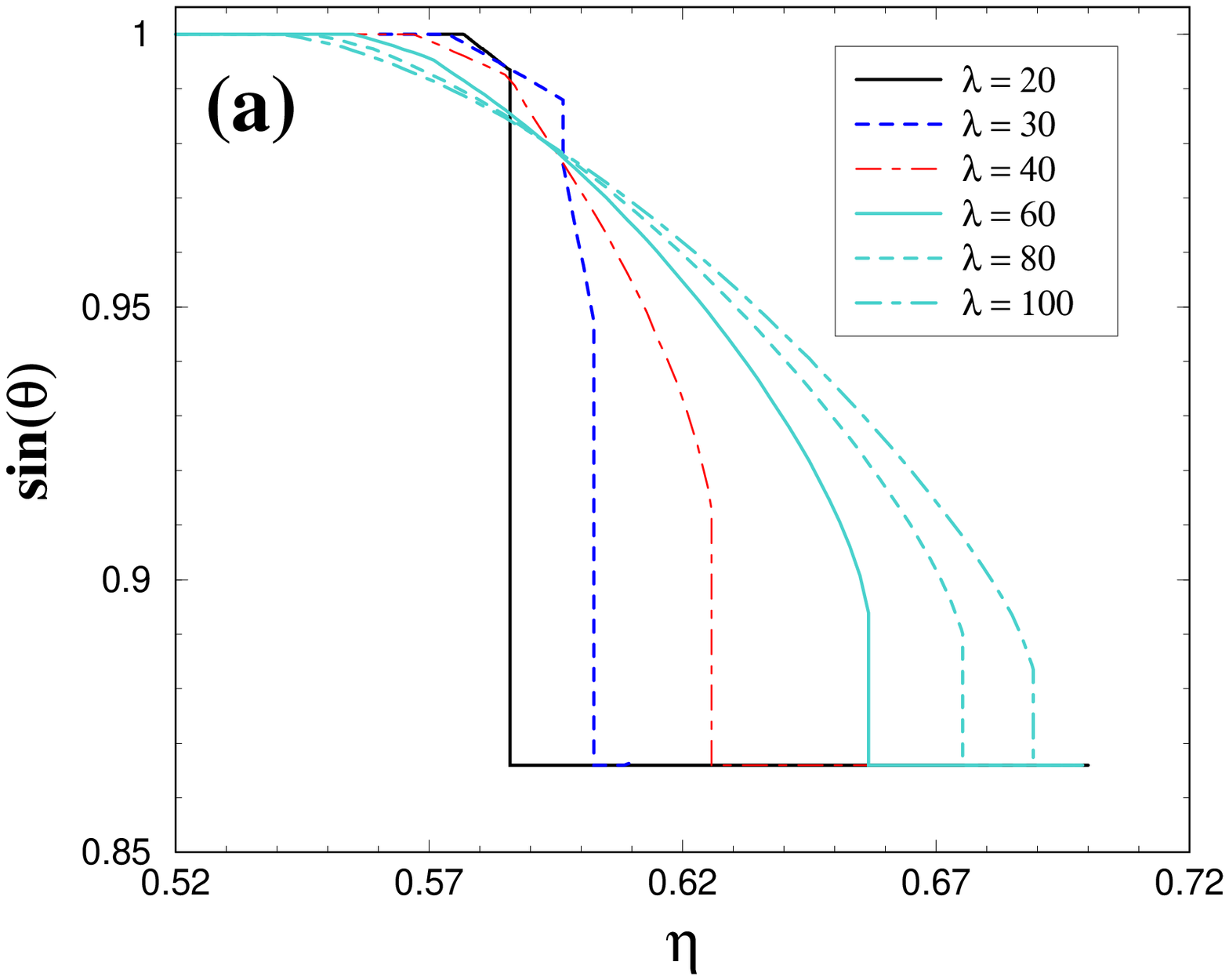}
\includegraphics[width = 8.0 cm]{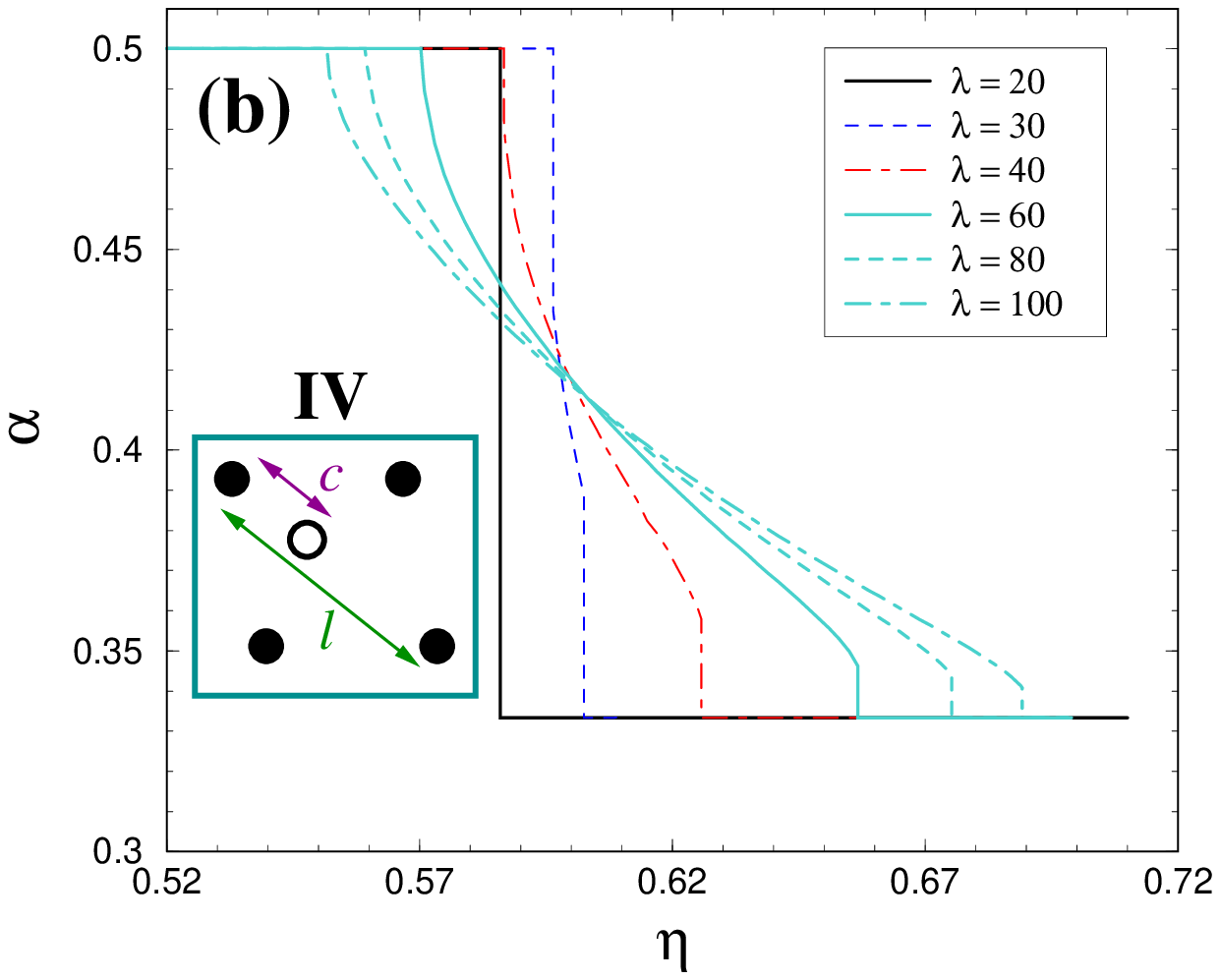}
\caption{
(a) Sine of the angle $\theta$ and (b) the relative shift parameter $\alpha = c/\ell$ 
(with $\ell = |{\bf a}_1 + {\bf a}_2|$)
versus density regarding the 
$\text{III} \rightarrow \text{IV} \rightarrow \text{V}$ transition scenario 
for different $\lambda$.
The insets show the lattice geometry of phase IV.
}
\label{fig.rhombic}
\end{figure}
%

Novel effects are observed for the $\text{III} \rightarrow \text{IV} \rightarrow \text{V}$
transitions. First, for small $\lambda$ the $\text{III} \rightarrow \text{V}$ transition
proceeds via a IVA phase, the former being second and the latter first-order.
For $\lambda \approx 8$, however, there is a strong first-order transition
directly from $\text{III}$ to $\text{V}$ with a large density jump as 
determined by Maxwell's construction \cite{note_maxwell}. For even higher screening,
$\lambda \gtrsim 30$, the $\text{III} \rightarrow \text{V}$ transition
happens via the cascade $\text{III} \rightarrow \text{VIA}\rightarrow \text{IVB}\rightarrow \text{V}$.
The stability range of the IVA phase becomes smaller for increasing $\lambda$
shrinking to zero in the hard sphere limit.
Details of the $\text{III} \rightarrow \text{V}$ transition scenario can be detected
via the order parameters $\sin \theta$ and $\alpha$ of the lattice minimizing the
total potential energy at prescribed  density $\eta$.
Plotting  $\sin \theta$ and $\alpha$ versus $\eta$ reveals the order of the transitions
(see Fig. \ref{fig.rhombic}):
A cusp, which is found for the $\text{III} \rightarrow \text{IVA}$ transformation, implies
a  second-order transition.
All other transitions are first-order as signaled by discontinuous jumps in at least
one of these order parameters.
The corresponding
coexistence density gap is not shown in Fig. \ref{fig.rhombic} but included
in Fig. \ref{Fig.diagram}(a). Across the $\text{IVA} \rightarrow \text{IVB}$ transition
the order parameter jump is small yielding a tiny density gap which
can not be resolved in Fig. \ref{Fig.diagram}(a).

Our most striking result is  {\it reentrant behavior} of the IVA phase at fixed density 
upon varying $\lambda$ as indicated in Fig. \ref{Fig.diagram} by the 
vertical arrow.
For $ 0.5 < \eta < 0.525 $, there is reentrance of the VIA phase via the III phase.
The full sequence over the whole range of $\lambda$ is
$\text{IVA}  \rightarrow \text{III} \rightarrow \text{IVA}\rightarrow \text{IVB}$.
For $ 0.530 < \eta < 0.536 $ there is even a {\it double
reentrant behavior} of the VIA phase 
via the sequence $\text{IVA} \rightarrow \text{V} \rightarrow \text{IVA} \rightarrow \text{III}  
\rightarrow \text{IVA}\rightarrow \text{IVB}$. This rich scenario is due to a subtle interplay of
the range of the interaction in conjunction  with the different bilayer lattice structures.
Finally, at {\it finite temperatures} $T$, we performed
extensive Monte-Carlo computer simulations with 800  particles in a rectangular-shaped box
periodically-repeated in $x$ and $y$ direction and with hard walls of distance $D$ in $z$-direction
allowing fluctuating $z$-positions of the particles (``buckling'').
For fixed $\eta=0.533$,
we investigated four states at $\lambda=0.5$, 3.0, 8.2, 10 
[see the four diamonds along the arrow in Fig. \ref{Fig.diagram}(b)] 
for different $T$ up to melting. The melting point is detected via a modified 
Lindemann-type criterion involving differences of mean-square displacements 
of nearest-neighbors \cite{Peeters}. We confirm that the reentrant behavior 
is stable with respect to increasing $T$ up to melting.

In summary, we have calculated the full phase diagram for a Yukawa bilayer
at zero temperature by lattice sum minimizations. 
A competition between three length scales,
namely the bilayer distance $D$, the averaged particle distance $\rho^{-1/2}$,
and the range $1/\kappa$ of the interaction, induces a rich phase behavior which is different 
from a simple interpolation of the extreme limits of the confined plasma and the hard sphere system.
We predict a coexistence of two different rhombic phases at
finite screening and a single and double  reentrant scenario for one of the rhombic phases 
for varied ``softness'' of the interaction. These effects are in principle
detectable in real-space experiments of charged colloidal
suspensions confined between plates and in layers of dusty plasmas by
tuning the screening strength via the microion concentration.
The reentrant effect as obtained here in equilibrium should
also manifest itself as an interesting fingerprint in nonequilibrium
situations. For example, bilayer crystal nucleation and growth could
be greatly stimulated via structures which are energetically close to the stable ones 
\cite{Palberg_review}.
Soft particle interactions different from the Yukawa type of Eq.~\eqref{Eq.Yukawa}, as
e.g.\ inverse power potentials where $V(r) \propto r^{-n}$, will
lead to similar reentrant effects as long as the softness of the potential
(e.g. the exponent $n$) is varied. 
Different realizations of soft
interactions occur in sterically-stabilized colloids, in spherical block-copolymer micelles
and in star polymers and dendrimers, 
where the softness of the effective interaction can be tuned by
the length and grafting density of the polymer chains or the solvent quality
\cite{Likos_PRL}.
Hence the reentrant scenario should also occur in foam films containing polymer bilayers 
\cite{kolaric}.
Finally, for a general external potential confining the particles
to layers, the  bilayer distance $D$ is not prescribed but the system will
minimize its total energy realizing an optimal $D$. In this case, 
second-order phase transitions will still be described in terms of scaled parameters.
This implies a {\it universal} behavior of
our bilayer phase diagram. In a general external potential, however, the system has 
the additional possibility to split into tri- and higher order multilayers.
This can happen either discontinuously or continuous via merging prism phases.
Details have to be explored in future studies.

\acknowledgments
We thank C. N. Likos and M. Schmidt for helpful discussions,
and the DFG (SFB TR6) for financial support.



\begin{thebibliography}{10}
\bibitem{note_peeters} For moderate screening strength $\lambda$,
the bilayer phase diagram (obtained by Monte Carlo simulations) was 
already reported in Ref. \cite{Peeters} but without
the reentrance effect, the density jump across
first-order transitions and the stable IVB phase.
\bibitem{point_particle} We neglect the non-overlap condition $r> 2R$ in Eq. (1)
since charged particles are typically not at contact.
\end{thebibliography}
\end{document}